\documentclass[letter]{aa} 

\usepackage{graphicx}
\usepackage{xcolor}
\usepackage{ulem}
\usepackage{txfonts}
\begin{document} 

\title{The behaviour of lithium at high metallicity in the Milky Way}
\subtitle{Selection effects in the samples and the possible role of atomic diffusion}

\authorrunning{C.Charbonnel et al.}
   
   \titlerunning{The behaviour of lithium at high metallicity in the Milky Way}

   \author{C. Charbonnel
          \inst{1,2}, 
          S. Borisov \inst{1,3},
          P. de Laverny \inst{4},
          N. Prantzos \inst{5}
          }
          
   \institute{Department of Astronomy, University of Geneva, Chemin Pegasi
   51, 1290 Versoix, Switzerland\\
              \email{Corinne.Charbonnel@unige.ch}
         \and IRAP, CNRS UMR 5277 \& Université de Toulouse, 14 avenue Edouard Belin, 31400 Toulouse, France
          \and Sternberg Astronomical Institute, M.V. Lomonosov Moscow State University, Universitetsky prospect 13, Moscow, 119234, Russia
         \and
         Université Côte d'Azur, Observatoire de la Côte d'Azur, CNRS, Laboratoire Lagrange, Nice, France
         \and
         Institut d’Astrophysique de Paris, UMR7095 CNRS, Sorbonne Université, 98bis Bd. Arago, 75104 Paris, France
             }

   \date{}

  \abstract
   {}
   {We revisit large spectroscopic data sets for field stars from the literature to derive the upper Li envelope in the high metallicity regime in our Galaxy.}
   {We take advantage of Gaia EDR3 data and state-of-the-art stellar models to precisely determine the position of the sample dwarf stars in the Hertzsprung-Russell diagram.}
   {The highest Li abundances are found in field metal-rich warm dwarfs from the GALAH survey, located on the hot side of the Li-dip. Their mean Li value agrees with what was recently derived for warm dwarfs in metal-rich clusters, pointing towards a continuous increase of Li up to super-solar metallicity. However, if only cool dwarfs are considered in GALAH, as done in the other literature surveys, it is found that the upper Li envelope decreases at super-solar metallicities, blurring the actual Li evolution picture. 
   We confirm the suggestion 
   that field and open cluster surveys that found opposite Li behaviour in the high metallicity regime do not sample the same types of stars: The first ones, with the exception of GALAH, miss warm dwarfs that can potentially preserve their original Li content.} 
  {Although we can discard the bending of the Li upper envelope at high metallicity derived from the analysis of cool star samples, we still need to evaluate the effects of atomic diffusion on warm, metal-rich early-F and late-A type dwarfs before deriving the actual Li abundance at high metallicity.
  }

   \keywords{stars: abundances - Galaxy: abundances - Galaxy: evolution}

   \maketitle
%

\section{Introduction}
\label{Section:Introduction}
Lithium-7 (Li) is the only element that has been produced in three extremely different astrophysical sites: in the very early universe during the Big Bang nucleosynthesis episode, in interstellar matter through cosmic ray spallation reactions and, mostly, in stars  of still undetermined type \citep[red giants, asymptotic giant branch stars, novae, and core collapse supernovae;][]{Matteucci1995,Travaglio2001,Romano2001,Prantzos2012b}. 
Quantitative predictions of Li evolution in different regions of 
the Milky Way hence remain challenging. 
It is generally expected, however, that the Galactic Li content is globally increasing with time (or metallicity), with variations depending on the regions considered (thin and thick discs, the bulge, and halo). This view is well supported by the observed upper envelope of Li abundances in low-mass stars (LMS) as a function of [Fe/H], which increases up to solar metallicity \citep[e.g.][]{Rebolo1988,BensbyLind2018}.

It thus came as a surprise when high-resolution spectroscopic data sets
such as AMBRE \citep{deLaverny2013,Guiglion2016} 
and others 
\citep{Delgado2015,Fu2018,Bensby2020,Stonkute2020} focusing on cool field main sequence (MS) stars indicated that Li may be decreasing at super-solar metallicities.
If this finding were confirmed, it would be the first time that an element other than H displays a decrease with metallicity \citep{Prantzos2017}. A few ideas were put forward to cope with this unprecedented situation \citep{Fu2018,Guiglion2019,Grisoni2019}, although \cite{Prantzos2017} had warned that the results deduced from AMBRE would require further analysis and additional observations before establishing such far reaching conclusions.
This finding has indeed been challenged recently, 
by observations of warm MS stars in metal-rich open clusters \citep[Gaia-ESO survey,][]{Randich2020} and in the field \citep[GALAH survey,][]{Gao2020}, which show high Li abundances above the meteoritic value, in line with Galactic chemical evolution expectations. 

One difficulty to determine the upper Li envelope as a function of metallicity as required to constrain Galactic chemical evolution models is that stars rarely exhibit in their photosphere the Li content they inherited from their protostellar cloud. 
This is due to the fragility of this light element, which is destroyed by proton captures at $\sim$2.5~MK in stellar interiors. 
There is plentiful observational evidence of surface Li depletion in FGK dwarfs along their early evolution, starting with the Sun 
whose photospheric Li abundance is $\sim$ 195 times lower than in meteorites \citep[][]{Wang2021}. 
A plethora of theoretical studies have investigated the internal transport processes that can lead to Li depletion in LMS and explain its dependence with spectral type, making this element one of the key constraints for modelling of LMS beyond standard evolution theory \citep[e.g.][and references therein]{Deliyannisetal2000IAUS,Dumontetal2021}.

In this paper, we do not discuss the various possible theoretical explanations for Li depletion or preservation. 
Instead, we focus on the observed Li behaviour in LMS of various types to pinpoint which ones are more relevant to constrain Galactic evolution models in the high metallicity regime (\S~\ref{Section:PMSandFtypestars}). We take advantage of the recent Gaia~EDR3 \citep{gaia_edr3} and of a large grid of stellar models to accurately determine the position in the Hertzsprung-Russell diagram (HRD) of stars from different surveys and to identify those that could have preserved their original Li at solar and super-solar metallicities. 
Using very strict selection criteria for the stellar parameters, parallaxes, and abundance determination, we derive the corresponding Li upper envelope (\S~\ref{Section:PMSandFtypestarsinSurveys}). 
We discuss the potential role of atomic diffusion in the  abundance patterns observed in super-solar metallicity warm dwarfs and conclude (\S~\ref{Section:Conclusions}).
 
\section{Looking for stars that might preserve their original surface Li} 
\label{Section:PMSandFtypestars}

Drawing the upper Li envelope as a function of [Fe/H] requires finding stars that exhibit Li abundances as close as possible to their original composition. 
Pre-main sequence (PMS) stars could be thought of as the best targets. However, observations in star forming regions, moving groups and associations, and very young open clusters have shown that PMS stars with masses lower than $\sim$ 0.9 -- 1~M$_{\odot}$ (mass limit estimate at solar metallicity; e.g. 
\citealt{Bodenheimer1965}) 
exhibit Li abundances that can be notably lower than the meteoritic value. 
Only more massive stars are seen to arrive on the zero age main sequence (ZAMS) without having undergone significant PMS Li depletion
\citep[][]{Magazzu1992,GarciaLopez1994,Soderblom1999,
James2006,Sestito2008,Balachandran2011,Messina2016,Jeffries2021}. 
Main sequence Li depletion has also long been evidenced in FGK dwarfs. It is
increasing with age and with decreasing stellar mass, as illustrated unambiguously by observations in open clusters \citep[e.g.][]{DuncanJones1983,Cayrel1984,Sestito2005,Anthony-Twarog2018} 
and in solar-like stars with reliable age estimates 
\citep[e.g.][and references therein]{Carlos2019,Melendez2020}.

The highest Li abundances found in open clusters with ages up to $\sim$ 1.5 -- 2~Gyrs are actually always exhibited by  
early type dwarf stars hotter than the so-called Li-dip
 (a drop-off in the Li content 
 centred around 6700~K that appears in all open clusters older than $\sim$~200~Mrs and in field MS stars; e.g. 
  \citealt[][]{Wallerstein1965,BoesgaardTripicco1986}). 
 These warm stars (T$_\mathrm{eff} \geq 6800$~K) show relatively constant Li abundances close to the cosmic value, indicating that they have undergone no or minimal surface Li depletion. 
This has been observed in open clusters and field stars with metallicities close to solar where the Li abundance of stars on the hot side of the Li-dip 
can reach the meteoritic value \citep[][]{HobbsPilachowski1986M67,HobbsPilachowski1986,BoesgaardTripicco1986field,BoesgaardTripicco1987,Boesgaardetal1988,BurkhartCoupry1989,Pasquini2001,Burkhart2005,Anthony-Twarog2021}, 
and in more metal-rich open clusters where even 
higher Li values (A(Li)$>$3.4) have been derived in 
the same effective temperature domain \citep[][]{Randich2020}. 
Some exceptions deviate from the mean, namely AmFm-type stars \citep{Burkhart2005,BurkhartCoupry1989,BurkhartCoupry2000} which are slow rotators that present Li deficiency (by about a factor of 3) compared to normal early A and late F type stars as well as other abundance anomalies due to atomic diffusion (gravitational settling and radiative levitation; discussion in  \S~\ref{Section:Conclusions}).

In summary, the best targets to  
observe the highest Li abundances and potentially build the Li upper envelope at solar metallicity and above, are either PMS or ZAMS stars with masses higher than $\sim$ 0.9~M$_{\odot}$ (at solar metallicity), or early-F type MS stars with effective temperatures higher than $\sim 6800~K$ (i.e., masses higher than $\sim$ 1.5~M$_{\odot}$ at solar metallicity) and with relatively high rotation rates to counteract atomic diffusion. 
Such objects will be absent in the metal-poor regime where the age of the sampled populations is larger than the duration of the PMS   
or the MS lifetime of early-F type stars ($\sim$ 1.8 to 2.9~Gyr for 1.5~M$_{\odot}$ models presented in this work with [Fe/H] between -0.23 and +0.5).   
Additionally, both groups of stars can be easily missed, or excluded, when building a volume-limited observational sample with completeness and selection function  often difficult to estimate accurately.
Technically, the Li resonance line at 6708~\AA~becomes weaker and weaker with an increasing effective temperature,
requiring high signal-to-noise spectra and optimised normalisation procedures. 
Moreover, relatively high rotation rates broaden the lines
and lead them to be even closer to the continuum level.
Such cases are thus more difficult to treat by automatic procedures. 

\section{Looking for PMS and early-F type stars in field star surveys at solar metallicity and above}
\label{Section:PMSandFtypestarsinSurveys}

\subsection{Selection criteria} 
\label{subsection:sampleselection}

We revisit the samples where a decline in the upper Li envelope at super solar metallicity has been found, putting more emphasis on AMBRE \citep{deLaverny2013,Guiglion2016} 
and briefly discussing the others \citep{Delgado2015,Fu2018,Bensby2020,Stonkute2020}. We also investigate the new release of the GALAH survey 
\citep{Buder2020,Gao2020}. In all cases we focus on the domain in [Fe/H] between -0.2 and +0.4~dex that we split into four bins of 0.15~dex. Bin~1 is sub-solar ($-0.20<$[Fe/H]$\leq-0.05$~dex), bin~2 is centred on solar metallicity ($-0.05<$[Fe/H]$\leq+0.10$~dex), and bins~3 and 4 are for the super-solar regimes ($+0.10<$[Fe/H]$\leq+0.25$ and $+0.25<$[Fe/H]$\leq+0.40$~dex,  respectively).  

We relied on the stellar parameters and abundances published in the original papers (T$_\mathrm{eff}$, log~$g$, [Fe/H], and A(Li)$_{\rm{LTE}}$), and kept only stars with log~$g\geq 4$. Since the GALAH data contain lithium abundance as [Li/Fe], we adopted A(Li)=[Fe/H]+[Li/Fe]+A(Li)$_{\odot}$ with A(Li)$_{\odot}$=0.96~dex \citep{Wang2021}. We applied non-local thermodynamic equilibrium (NLTE) corrections according to \citet{Wang2021} to the published AMBRE LTE Li abundances; this NLTE correction was already applied in the GALAH sample.
We applied the quality criteria provided in the reference papers to select the stars with the most reliable abundances and stellar parameters. In GALAH, we selected only stars with \texttt{flag\_li\_fe}=0 (quality flag on [Li/Fe], see \citealt{Buder2020} for details), \texttt{flag\_sp}=0 (stellar parameters quality flag), and  S/N per pixel $\geq$ 30. 

We kept only those stars with Gaia~EDR3 parallax uncertainties lower than 5$\%$ and RUWE$<$1.4. We used G-band photometry from Gaia~EDR3 and distances $d$ 
from \citet[][excluding possible less-quality parallaxes thanks to the RUWE parameter]{bailer-jones2020} to determine the luminosity $L$ of the sample stars, defined as $L=10^{0.4(M^{bol}_{\odot}-(G+5-5\mathrm{log}(d)+BC_G-A_G))}$ in units of $L_{\odot}$, where $M^{bol}_{\odot}=4.74$~mag 
(IAU resolution 2015, \citealt{prsa2016}). The bolometric correction $BC_G$ was computed according to \citet{andrae2018} 
 adopting the effective temperatures provided by the ground-based surveys. The $A_G$ is the G-band extinction from Gaia~DR2, 
 which is consistent with the DR2 colours adopted for deriving DR2 $BC_G$ and T$_\mathrm{eff}$.
Error on luminosity (typically $\sim$ 10~$\%$) is dominated by the one on $A_G$ (typically $\sim$ 0.11~mag), which is small enough to not affect our conclusions. Finally,  
 since a degeneracy could exist between the Gaia~DR2 derived T$_\mathrm{eff}$ and extinction, we kept only stars having consistent effective temperatures: $|\mathrm{T_{eff}^{Gaia}}-\mathrm{T_{eff}^{GALAH}}|\leq200$~K. With these selection criteria, for [Fe/H]>-0.2~dex, we have 962 stars in AMBRE, and 21130 stars in GALAH.

To distinguish PMS stars, we made a cross-match of the samples with the SIMBAD database \citep{simbad}.  
We looked for stars of the following types: T~Tauri star, young stellar object`, and variable star of Orion type``. As can be seen from Fig.~\ref{hrdsAMBREGALAH}, the GALAH sample does not contain PMS stars, in contrast to the AMBRE sample. This is because 
we selected only stars being flagged by GALAH
with \texttt{flag\_sp}=0 (\texttt{flag\_sp}=64 refers to both binary and PMS stars). 

\subsection{Positions of the sample stars in the HRD and 
determination of the Li upper envelope}
\label{subsection:HRD}

\begin{figure}[t]
  	 \center
   	 \includegraphics[width=1\linewidth]{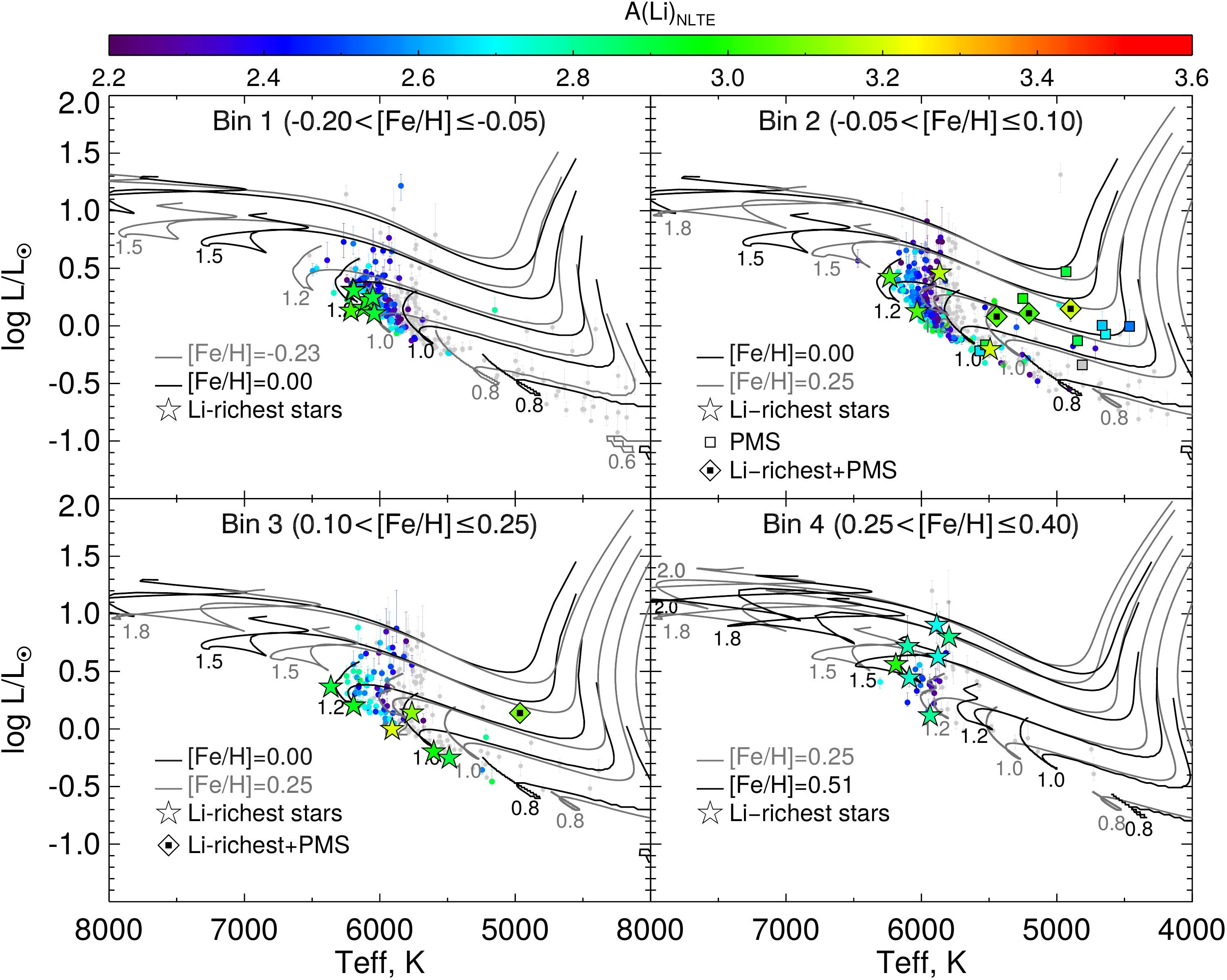}
   	 \includegraphics[width=1\linewidth]{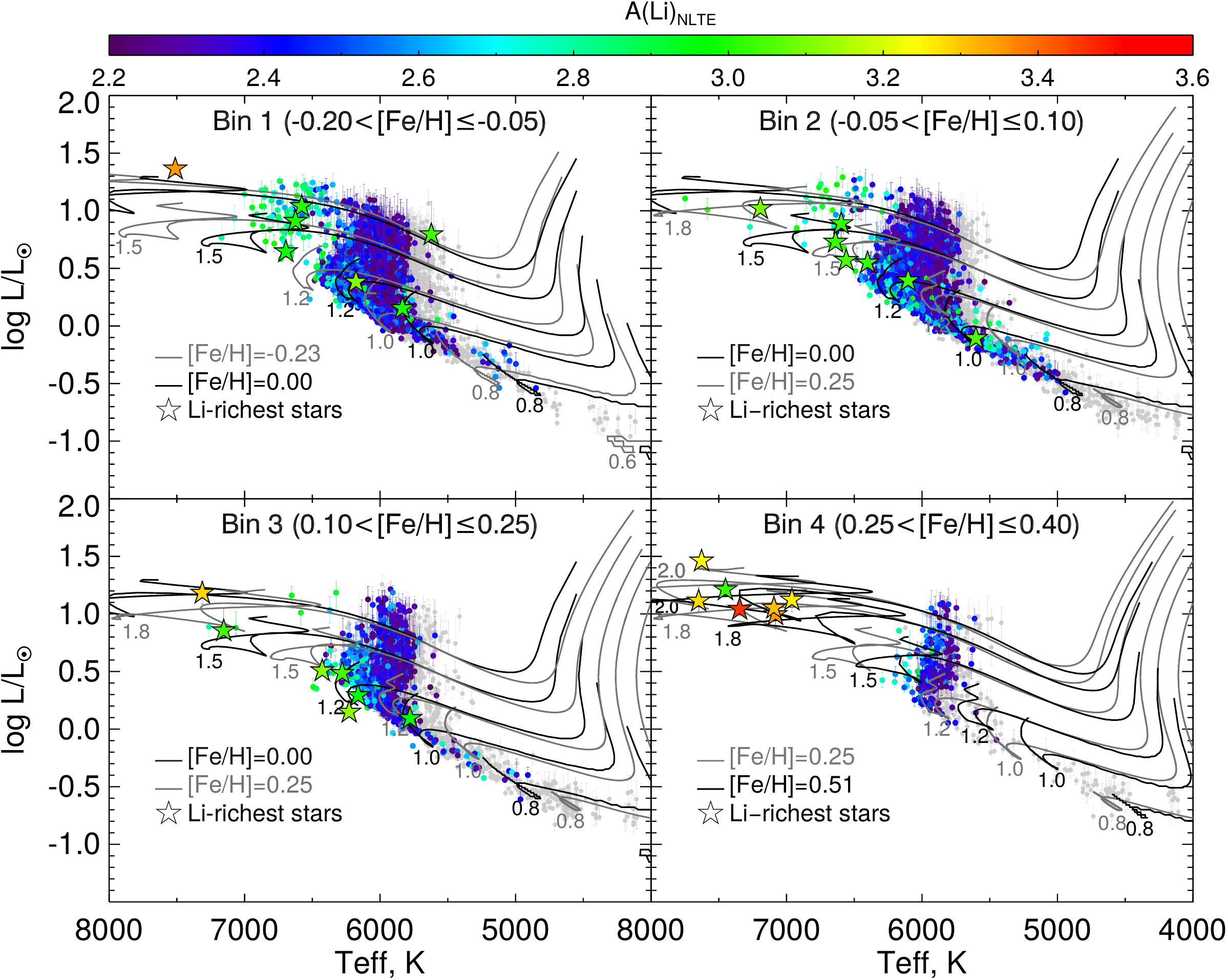}
   	 \caption{Position of the AMBRE and GALAH sample
   	  stars (top and bottom respectively) in the different metallicity bins, with the Li abundance colour-coded (stars with A(Li) lower than 2.2 are all in grey). In each panel, the seven stars with the highest lithium abundances of the corresponding metallicity bin are shown with star symbols (when MS) or diamonds (when PMS). We indicate other
   	  stars that were found to be PMS (squares). Evolution tracks from \citet{Lagarde2017} are shown (PMS and MS) for [Fe/H] values, covering the respective observational ranges.}
   	 \label{hrdsAMBREGALAH}
 \end{figure}
 
Figure~\ref{hrdsAMBREGALAH} shows the position of individual stars in the four metallicity bins for AMBRE and GALAH samples, respectively. NLTE Li abundances are colour-coded, and the seven Li richest stars in each individual metallicity bin are highlighted, as well as the known PMS stars. PMS and MS evolutionary tracks from \citet{Lagarde2017} completed by tracks at [Fe/H]=+0.25 computed 
with the same code and input physics are also shown. 

We immediately see that there are no late-A and early-F type metal-rich dwarf star with T$_\mathrm{eff}>6800$~K in the 
AMBRE sample (see also Fig.~3 of \citealt{Guiglion2016}), and that in the most metal-rich bin the seven Li-richest stars are cool stars lying close to other stars which show very strong Li depletion. 
The lack of warm stars is inherent to the AMBRE/Li catalogue. 
First, the AMBRE
sample consists in a wealth of ESO archived spectra that
have been collected for various scientific goals. The AMBRE
selection function is thus very complex and could reflect
a possible lack of early-F type stars observed with ESO high-resolution spectrographs.
Moreoever, within AMBRE, the stars are parametrised thanks to
automatised pipelines that rely on pre-computed grids of non-rotating FGKM-type stellar spectra. Such a methodology 
thus a priori rejects non-rotating stars hotter than $\sim$7,500~K and any cooler ones having rotation rates larger than $\sim 15$~km/s \citep[][]{Worley2012},
leading to a final sample biased towards cool and slowly rotating stars.
We did the same analysis for the samples of \citet{Delgado2015}, \citet{Fu2018},
and \citet{Stonkute2020}\footnote{\citet{Bensby2020} focus on the Galactic bulge and have only 3 stars with well-determined Li, logg>=4 and -0.2<Fe/H<=0.4, and they are all cool G-type stars (T$_\mathrm{eff}$=6130~K, 5732~K, and 5947~K).}. As in AMBRE, none of the sample stars used for the derivation of the Li upper envelope in the high metallicity regime lie on the hot side of the Li dip. 

\begin{figure}[t]
  	 \center
   	 \includegraphics[width=1\linewidth]{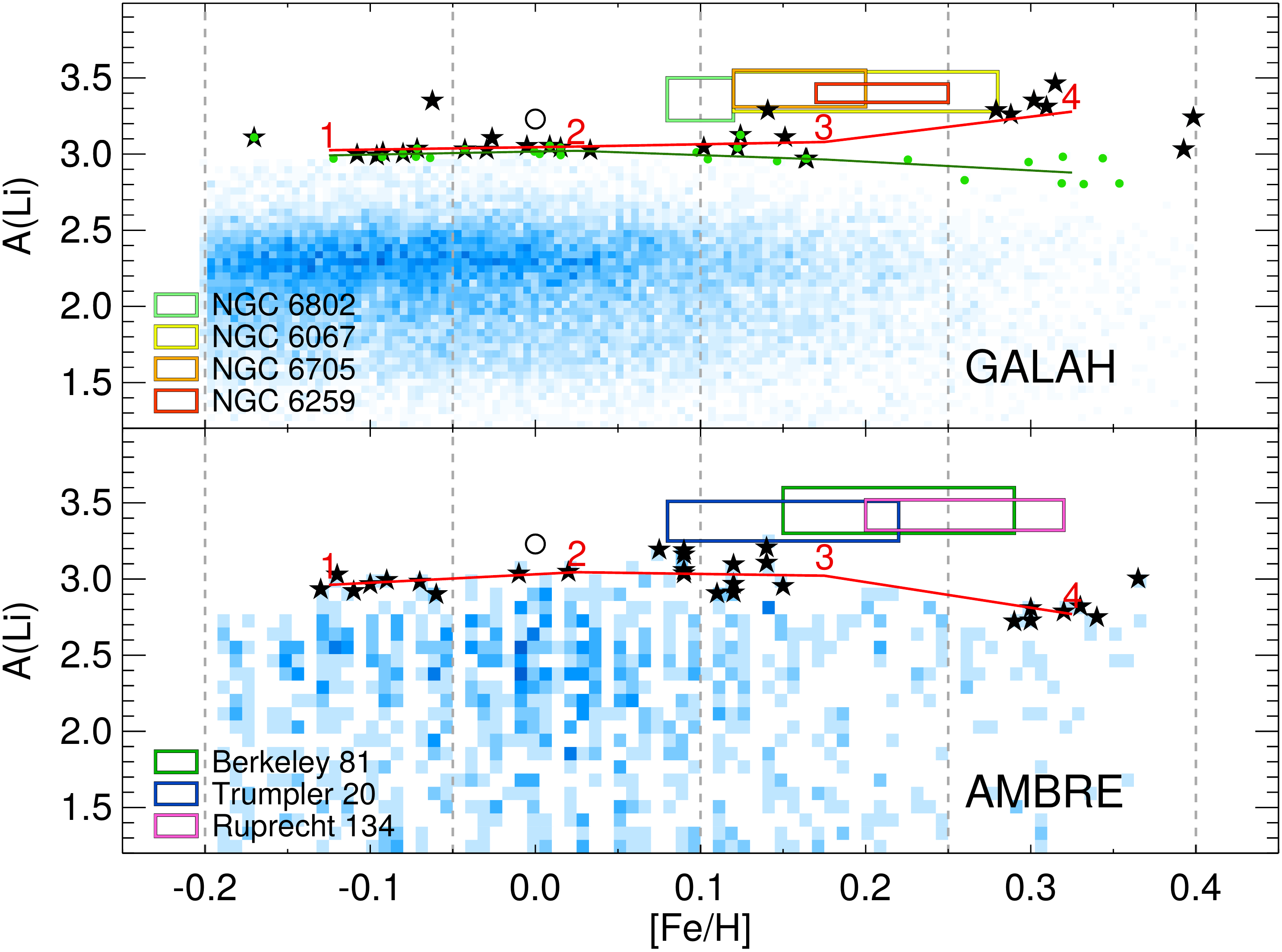}
   	 \caption{Number density plot showing lithium abundance A(Li) of the GALAH~DR3 (upper panel) and AMBRE (bottom panel) stars versus [Fe/H] (colour-coded number densities have different scales on the panels). The  upper Li-envelope (red line) connects the mean A(Li) of the seven most Li-rich stars (black stars) in each bin. The dark green line connects the mean A(Li) obtained with GALAH data when using only cool stars from the same T$_\mathrm{eff}$ range as in the AMBRE sample (green points). Coloured boxes represent super-solar metallicity open clusters with sizes of $1\sigma$ confidential intervals on [Fe/H] and A(Li)$_{\rm max}$ from \citet{Randich2020}. In all the clusters shown, the Li data 
   	    were obtained for early-F type stars from the hot side of the Li dip. The open circle indicates the meteoritic value.} 
   	 \label{LiFeH_AMBRE_GALAH}
 \end{figure}

In contrast, early-F and late-A type stars with T$_\mathrm{eff}>6800$~K are present in the four metallicity bins of the GALAH sample (Fig.~\ref{hrdsAMBREGALAH}), and they always exhibit the highest Li abundances in the respective bins, which is in agreement with what is found in open clusters. The average Li value we get for the seven Li-richest stars in this T$_\mathrm{eff}$ domain for the highest metallicity bin is 3.28~dex, with the Li richest star having 3.46~dex. As can be seen in Fig.~\ref{LiFeH_AMBRE_GALAH}, this agrees perfectly with the  average maximum A(Li) values for the supersolar metallicity open clusters by \citet{Randich2020} for which the authors selected upper MS stars on the warm side of the Li dip (above 6500~K actually). 
Their Li abundances were not corrected for NLTE effects, but the authors argue that these corrections should be significantly lower than +0.1~dex. 
The A(Li) trend with [Fe/H] they obtain is in very good agreement with the one we derived when we used the similarly Li-rich stars from the hot side of the dip in the GALAH sample. 

Finally, we also show in Fig.~\ref{LiFeH_AMBRE_GALAH} the average Li value of the seven most Li-rich stars of GALAH in the reduced cool T$_\mathrm{eff}$ range of AMBRE for each of the four bins. We then find a very good agreement with the trend obtained from AMBRE. 
This confirms the suggestion of \citet{Randich2020} to explain the discrepancy between their analysis that contains such warm stars and previous works for the high metallicity bins.
We clearly show the reason why the upper envelope 
is found to decrease in the high metallicity regime for these surveys is because of secular Li depletion in cool dwarfs. 

 \begin{figure}[t]
  	 \center
   	 \includegraphics[width=1\linewidth]{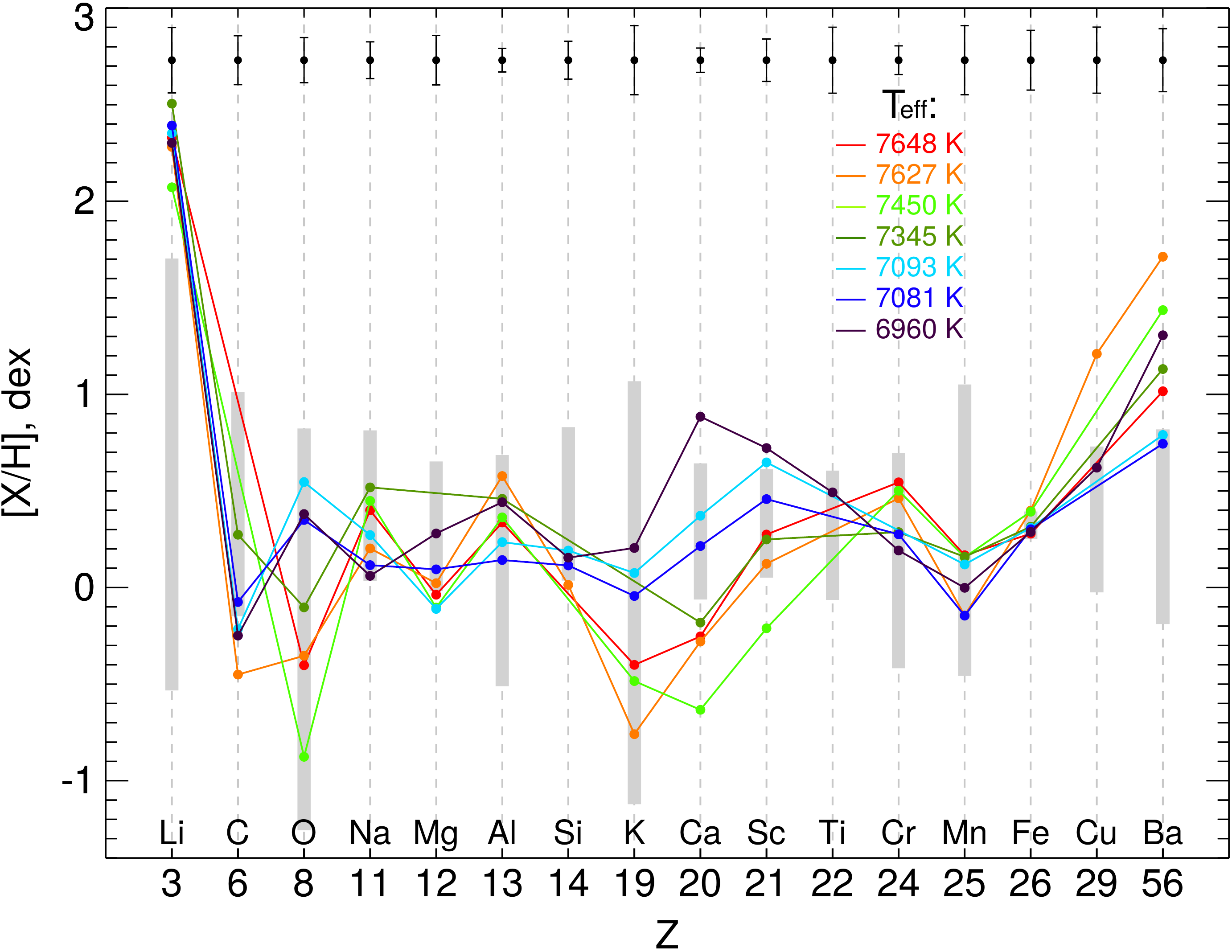}
   	 \caption{Surface abundance of different elements (normalised to solar photospheric abundances) in the Li-richest stars of GALAH bin~4, colour-coded according to their Teff. Grey boxes represent the corresponding abundance ranges of G-type stars from bin~4. Error bars at the top show the mean [X/H] errors.}
   	 \label{elementsGALAH}
 \end{figure}

\section{Toward a reliable conclusion on the Li trend in the high metallicity regime}
\label{Section:Conclusions}

Since we are mostly interested in the metal-rich regime, we focus our discussion on the seven Li-richest stars of GALAH bin~4 (+0.25 < [Fe/H] < +0.4) that have a T$_\mathrm{eff}$ between 6960 and 7650~K.
As already mentioned, deriving abundances in warm, possibly fast rotating, dwarf stars, is challenging. However, we consider only stars with high resolution, high S/N spectra; the S/N per resolution element range is between 122 and 965 for these seven Li-rich stars. 
Additionally, they are relatively slow rotators for their spectral types ($v\sin i$ between 16 and 52 km.sec$^{-1}$ according to GALAH~DR3), and they are supposedly single stars (\texttt{flag\_sp}=0). 
Specific tests based on simulated spectra of such rotating stars confirmed that the derived Li abundances are reliable for the considered high S/Ns.

On the other hand, stars with a T$_\mathrm{eff}$ higher than $\sim$ 6800~K may show superficial abundance anomalies due to atomic diffusion under the competitive action of gravitational settling and radiative acceleration, unless meridional circulation, turbulence, mass loss, or a combination of these processes partially compensate or eventually erase chemical separation effects \citep[e.g.][]{Michaudetal1983,VV1982,Turcotte1998}. 
The actual rotation threshold between normal A-type stars and chemically peculiar Am ones is, however, not firmly established observationally or theoretically (it varies in the literature between $\sim$ 60 and 120 km.sec$^{-1}$), 
with also binarity playing a possible role \citep[][]{Abt2000,VarenneMonier1999,Royeretal2014}. 

We thus looked  for possible signatures of chemical separation in the seven Li-richest stars of GALAH bin 4, using DR3 abundances \citep{Buder2020}. We checked  particular elements that are expected to be overabundant (Ti, Cr,  Mn, Fe, Ni, and Ba) and others that are expected to be underabundant (Li, C, O, Mg, Si, K, and Ca) in AmFm-type stars \citep{Preston1974,Richard2001,Talon2006,Vick2010} 
compared to normal A-type stars \citep{HillLandstreet1993,VarenneMonier1999,Adelman2000}.
The corresponding [X/H] values are shown in Fig.~\ref{elementsGALAH}, together with those of other elements that should not be significantly affected by atomic diffusion in these warm stars (Na, Al, and Sc). 
For each element, we also show 
the observed abundance range in G-type stars of bin~4, which exhibits important star-to-star abundance variations. Since the surface abundances of these cool stars are not affected by atomic diffusion, this reflects their initial chemical composition (except for Li, which is depleted, as is expected and described in Sect.\ref{Section:PMSandFtypestars}), hence the expected range of abundances that bin~4 stars could have inherited at birth.

Importantly, for most of the elements, the dispersion in abundances among the seven Li-rich stars is in good agreement with (and, for some elements, lower than) the spread in initial abundances depicted by G-type stars. 
The only elements for which the seven stars deviate from the bulk are Ca, Cu, and Ba, with one star also showing a slight underabundance in C.
The Ba overabundances observed in five of the seven stars are, however, lower than the minimum value 
assumed by \citet{Xiang2020} to characterise metal-rich AmFm stars; 
additionally, for this element, NLTE effects could potentially be important \citep{Mashonkina2020} and may blur the comparison with the abundances derived for cool stars. This analysis thus tends to indicate that the seven Li rich stars in GALAH bin~4 have relatively normal abundances of heavy elements.

The actual status (AF versus AmFm) of the warm metal-rich and Li-rich stars however deserves confirmation. Taylor-made evolution models with atomic diffusion, rotation, and mass loss are required in particular to make sure that their [Fe/H] values reflect the original composition they inherited at birth and do not result from chemical separation. It would also be useful to compare their heavy element abundance patterns with those of the warm stars in the metal-rich clusters analysed by \citet{Randich2020}, for which the original [Fe/H] can be obtained from the analysis of cool cluster members that do not undergo chemical separation.

Importantly, atomic diffusion could possibly decrease the surface Li abundance of these stars with respect to the original one, as observed in AmFm stars in open clusters \citep{BurkhartCoupry1989,BurkhartCoupry1997,BurkhartCoupry2000} and as predicted by models for solar metallicity stars \citep{Talon2006,Vick2010}.
In that case, the original Li abundance in metal-rich Galactic environments could be even higher than the mean value derived in this work and in \citet[][]{Randich2020}. Interestingly, the Li-richest star (A(Li)=3.46~dex) of GALAH bin~4 is also the fastest rotator, hence it is possibly the one with less Li depletion due to atomic diffusion.

We therefore conclude that the trend derived from cool MS stars can be discarded, because of Li depletion along their evolution. On the other hand, while we clearly show that the cosmic Li abundance has kept increasing in the Galaxy, the derivation of its actual value in the high metallicity regime deserves further analysis.

\begin{acknowledgements}
We are grateful to Nad\`ege Lagarde for providing part of the grid of stellar evolution models we use in this work and to J\'er\^ome Bouvier 
and Marc Audard 
for fruitful discussions on PMS stars. We are thankful to the anonymous referee for the important comments. This work was supported by the Swiss National Science Foundation (Project 200020-192039 PI C.C.). It has made use of the SIMBAD database, VizieR catalogue access tool (both operated at CDS, Strasbourg, France), and TOPCAT \citep{topcat}. This work has made use of data from the European Space Agency (ESA) mission Gaia, processed by the Gaia Data Processing and Analysis Consortium (DPAC). Funding for the DPAC has been provided by national institutions, in particular the institutions participating in the Gaia Multilateral Agreement.
\end{acknowledgements}

\bibliographystyle{aa}
\bibliography{Charbonnel_40873_Reference}

\end{document}